\documentclass[sn-nature]{sn-jnl}

\usepackage{graphicx}%
\usepackage{multirow}%
\usepackage{amsmath,amssymb,amsfonts}%
\usepackage{amsthm}%
\usepackage{mathrsfs}%
\usepackage[title]{appendix}%
\usepackage{xcolor}%
\usepackage{textcomp}%
\usepackage{manyfoot}%
\usepackage{booktabs}%
\usepackage{algorithm}%
\usepackage{algorithmicx}%
\usepackage{algpseudocode}%
\usepackage{listings}%

\usepackage{comment}
\newcounter{comments}
\definecolor{Red}{rgb}{0.8,0,0}
\definecolor{Green}{rgb}{0.2,0.6,0.2}

\newcommand{\beginsupplement}{%
        \setcounter{table}{0}
        \renewcommand{\thetable}{S\arabic{table}}%
        \setcounter{figure}{0}
        \renewcommand{\thefigure}{S\arabic{figure}}%
     }

\begin{document}

\title[Canalization reduces the nonlinearity of regulation in biological networks]{Canalization reduces the nonlinearity of regulation in biological networks}


\author*[1]{\fnm{Claus} \sur{Kadelka}}\email{ckadelka@iastate.edu}

\author[2]{\fnm{David} \sur{Murrugarra}}\email{murrugarra@uky.edu}

\affil*[1]{\orgdiv{Department of Mathematics}, \orgname{Iowa State University}, \orgaddress{\street{411 Morrill Rd}, \city{Ames}, \postcode{50011}, \state{IA}, \country{United States}}}

\affil[2]{\orgdiv{Department of Mathematics}, \orgname{University of Kentucky}, \orgaddress{\street{719 Patterson Office Tower}, \city{Lexington}, \postcode{40506}, \state{KY}, \country{United States}}}


\abstract{
Biological networks such as gene regulatory networks possess desirable properties. They are more robust and controllable than random networks. This motivates the search for structural and dynamical features that evolution has incorporated in biological networks. A recent meta-analysis of published, expert-curated Boolean biological network models has revealed several such features, often referred to as design principles. Among others, the biological networks are enriched for certain recurring network motifs, the dynamic update rules are more redundant, more biased and more canalizing than expected, and the dynamics of biological networks are better approximable by linear and lower-order approximations than those of comparable random networks. Since most of these features are interrelated, it is paramount to disentangle cause and effect, that is, to understand which features evolution actively selects for, and thus truly constitute evolutionary design principles. Here, we show that approximability is strongly dependent on the dynamical robustness of a network, and that increased canalization in biological networks can almost completely explain their recently postulated high approximability.

}

\keywords{Boolean networks, systems biology, nonlinear dynamics, canalization, stability, approximation}



\maketitle

\section{Introduction}\label{sec1}
Biological systems are frequently represented as networks, which describe the interactions between different biological entities such as genes, proteins or metabolites. For instance, gene regulatory networks (GRNs) describe how a collection of genes governs key processes within a cell. A \emph{static} biological network is completely described by a wiring diagram, which contains nodes (e.g., genes) and edges between nodes, which can be undirected (e.g., in protein-protein interaction networks), directed and even signed (e.g., in gene regulatory networks). Static networks are, however, insufficient to obtain accurate insights into the often complex, non-linear dynamics of biological networks~\cite{barrat2008dynamical}. \emph{Dynamic} biological networks possess the additional information how each node is regulated by the set of its regulators. Popular dynamic modeling frameworks include differential equation models and discrete models. While the former harbors the potential for quantitative predictions, it requires a substantial amount of data for an accurate inference of its many kinetic parameters. Therefore, many modelers prefer discrete models and their qualitative predictions. Boolean networks constitute the simplest type of discrete model. Here, each node takes on only two values and time is discretized as well. The two values can be interpreted as low and high concentration, unexpressed and expressed genes or proteins, etc. Particularly for GRNs, Boolean networks have become increasingly popular. Over 160 Boolean GRN models have been curated by experts in their respective fields - most over the course of the last twelve years~\cite{kadelka2024meta}. These models range in size from 3 to 302 nodes, and describe various processes in many species and kingdoms of life.

Over the last few decades, a number of interesting features of biological networks have been identified. At the structural, ``wiring diagram" level, biological networks are sparsely connected with an average degree of about $2.5$, and are enriched for certain network motifs such as coherent feed-forward loops and complex feedback loops, particularly those that contain many negative interactions~\cite{shen2002network,kadelka2024meta}. Dynamically, most biological networks operate at the critical edge between order and chaos~\cite{balleza2008critical,daniels2018criticality,kadelka2024meta}. For random $N-K$ Kauffman networks, it is well-established that the network dynamics are generally ordered whenever $2Kp(1-p)<1$ and chaotic whenever $2Kp(1-p)>1$; at $2Kp(1-p)=1$, a phase transition happens~\cite{luque2000lyapunov,shmulevich2004activities}. Here, $K$ is the average degree of the network, while $p$ describes the \emph{bias} of picking a one in the Boolean function's truth table; the unbiased case corresponds to $p=0.5$, and the absolute bias can be quantified by $2|0.5-p|\in[0,1]$, or alternatively by $1-4p(1-p)\in[0,1]$, with $0$ corresponding in both cases to the unbiased case. Networks with ordered dynamics typically possess few and short attractors, while chaotic dynamics are characterized by the presence of many long attractors~\cite{chandrasekhar2023stability}.

The dynamic update rules of Boolean biological network models are also remarkable. They are highly canalizing, redundant and biased~\cite{kadelka2024meta,harris2002model,gates2021effective}. 
\emph{Canalization} is a widely used term in biology. First coined by developmental geneticist Waddington in the 1940s~\cite{waddington1942canalization}, it refers to the tendency of developmental processes to follow particular trajectories, despite internal and external  perturbations~\cite{hallgrimsson2002canalization}. In other words, it refers to low variation in phenotypes despite potentially high variation in genotypes and the environment~\cite{flatt2005evolutionary}. 
Correspondingly, Kauffman introduced Boolean \emph{canalizing} functions as suitable update rules to describe the gene regulatory logic~\cite{kauffman1974large}. A canalizing function possesses a canalizing variable, which, when it receives its canalizing input, determines the output of the function, irrespective of all other inputs. If the subfunction which is evaluated when the canalizing variable does not receive its canalizing input is also canalizing, the function is 2-canalizing, etc~\cite{layne2012nested}. If all $n$ variables of a function become eventually canalizing, the function is $n$-canalizing, also known as \emph{nested canalizing}~\cite{kauffman2003random}. The number of variables which become eventually canalizing is known as the \emph{canalizing depth}~\cite{layne2012nested}. Every non-zero Boolean function possesses a unique standard monomial form, from which the canalizing depth and the number of variables in each ``layer" of canalization can be directly derived~\cite{he2016stratification,dimitrova2022revealing}. As the number of variables increases, canalizing and especially nested canalizing functions become increasingly rare~\cite{just2004number,li2013boolean,kadelka2017multistate}. It is therefore very surprising that almost all rules in published Boolean biological network models are canalizing and even nested canalizing~\cite{kadelka2024meta,harris2002model}.

Another recently discovered feature of biological Boolean network models is the high \emph{approximability} of their dynamics by linear and low-order continuous Taylor approximations of the Boolean update rules~\cite{manicka2023nonlinearity}. Here, the mean approximation error (MAE) is defined as follows: Each update rule of a given Boolean network is replaced by a continuous Taylor approximation of a defined order. The MAE describes the mean squared error between the long-term state of the Boolean network and the long-term state of the continuous approximation when starting from a random initial state (see Methods for details). Manicka et al found that biological networks were consistently more approximable (i.e., had lower MAE values) than random networks with the same wiring diagram (i.e., matching degree distribution) and matching update rule bias~\cite{manicka2023nonlinearity}.

Many of the described remarkable features of biological networks are interrelated and correlated. For instance, canalizing Boolean functions are on average more redundant and biased than random functions~\cite{kadelka2024meta}. In this paper, we show that the described increased approximability of biological networks can be fully explained by the abundance of canalization, which was not considered in~\cite{manicka2023nonlinearity}. We further show that the approximability of a Boolean network depends mostly on its dynamic regime, which in turn depends on its update rules (that is, average degree, bias and amount of canalization)~\cite{kadelka2024meta,manicka2022effective}. A network with ordered dynamics (i.e., few and short attractors) tends to possess much more approximbale dynamics than a network with chaotic dynamics.

\section*{Results and discussion}
\begin{figure}
    \centering
    \includegraphics[scale=1]{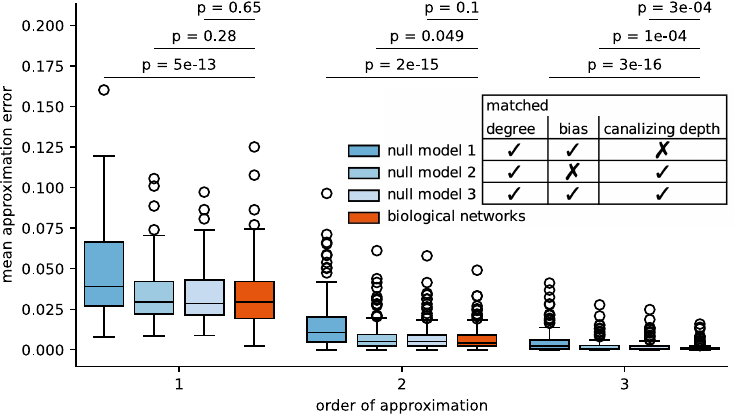}
    \caption{{\bf Canalization explains the high approximability of biological networks.}. The distribution of mean approximation errors is shown for the biological networks (orange) and three different types of random null networks (shades of blue), which match different characteristics (bias and/or canalizing depth) of the biological network. Each box depicts the interquartile range (IQR), each whisker extends to the most extreme value within 1.5 * IQR from the box, and each horizontal line within a box depicts the median. For a fixed approximation order (1-3, x-axis), differences between the MAE distribution of the biological and the random networks are assessed using the two-sided Wilcoxon signed-rank test. Fig.~\ref{fig:bio_networks_approximability_detailed} contains scatterplots showing the MAE values of all biological networks and their random null models.}
    \label{fig:bio_networks_approximability}
\end{figure}

\begin{figure}
    \centering
    \includegraphics[width=\textwidth]{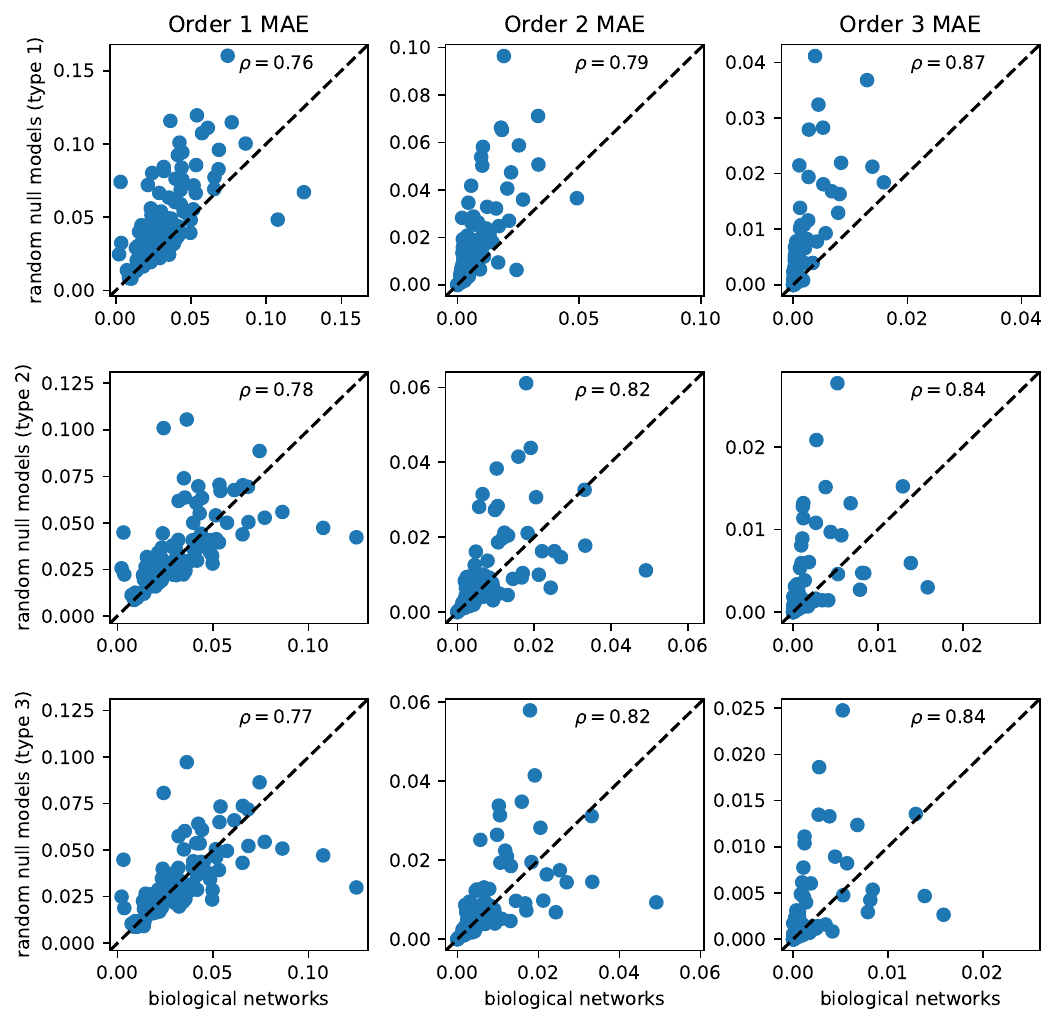}
    \caption{{\bf Mean approximation errors of biological networks and their random null models.} For a fixed approximation order (1-3, columns), differences between the MAE values of the 110 biological and the different random networks (rows) are shown, in addition to the Spearman correlation coefficient, $\rho$. A summary of this data is shown in Fig.~\ref{fig:bio_networks_approximability}.}
    \label{fig:bio_networks_approximability_detailed}
\end{figure}

To test the hypothesis that the increased canalization in biological networks explains their increased approximability, we compared the approximability of published expert-curated biological networks with several ensembles of random null models, similar to~\cite{manicka2023nonlinearity}. All random networks possessed the same wiring diagram as the respective biological network. The authors in~\citep{manicka2023nonlinearity} considered an ``unconstrained" null model, where each biological update rule was replaced by a non-constant random Boolean function (of the same degree), and a ``constrained" model (null model type 1 in this study), which additionally matched the bias of each biological update rule. Neither model accounted for the high degree of canalization in biological networks. We therefore considered two additional null models, one which matches the degree and canalizing depth of each biological update rule (null model type 2), and one which matches degree, canalizing depth and bias (null model type 3; see Methods for details). After excluding highly similar biological models and those with a maximal degree of eleven or more, we compared the approximability of 110 published expert-curated biological Boolean network models~\cite{kadelka2024meta} and the three different ensembles of null models. As in~\cite{manicka2023nonlinearity}, we found that random networks of type 1 were less approximable~(Fig.~\ref{fig:bio_networks_approximability}). However, random networks that accounted for the increased canalization (null models of type 2 and type 3) exhibited similar levels of approximability as the biological networks. Interestingly, the higher the order of employed approximation the more significant were the differences in the MAE distributions between biological and random networks~(Fig.~\ref{fig:bio_networks_approximability_detailed}). Third-order Taylor approximations recovered the dynamics of biological networks slightly better than those of random networks with matched degree, bias and canalizing depth.
Overall, these results show that the approximability of biological networks can be almost entirely explained by their high degree of canalization, measured by the canalizing depth.

However, a related question, which has implications for the control of Boolean networks~\cite{borriello2021basis},
remains: Why can the dynamics of biological networks be approximated so well by low-order and even linear continuous Taylor approximations? We hypothesized that the approximability of a Boolean network is strongly correlated with its dynamical robustness, which is typically measured by the average sensitivity~\cite{shmulevich2004activities} and Derrida values~\cite{Derrida1,Derrida2}. These metrics describe how a small perturbation affects the network over time. If the perturbation gets on average smaller after each node has been synchronously updated once, the system operates in the \emph{ordered} regime; if, on average, it increases in size, the system is in the \emph{chaotic} regime, and if it remains, on average, of similar size, the system exhibits \emph{criticality}.  All biological systems that have thus far been modeled as Boolean networks operate close to the critical edge between order and chaos~\cite{balleza2008critical,daniels2018criticality, kadelka2024meta}. 
This is likely because most update rules in biological networks are nested canalizing - in fact, biological networks are even particularly enriched for insensitive nested canalizing functions (NCFs)~\cite{kadelka2024meta} - and the expected average sensitivity of an NCF in any number of variables is 1. On the contrary, the average sensitivity of random Boolean functions with degree $k$ and bias $p$ is $2kp(1-p)$. That is, it increases as the number of inputs increases and decreases as the function becomes more biased (where $p=0.5$ corresponds to the unbiased case). Boolean networks governed by such random functions thus exhibit a phase transition at $2kp(1-p)=1$~\cite{luque2000lyapunov,shmulevich2004activities}.

\begin{figure}
    \centering
    \includegraphics[scale=0.85]{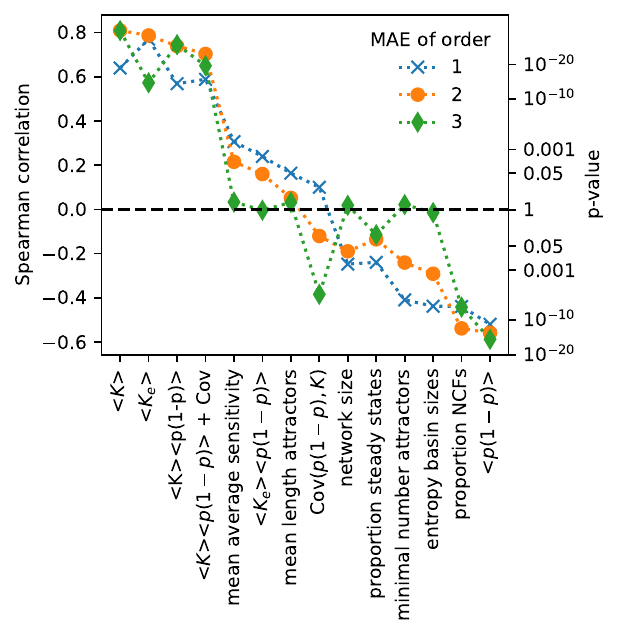}
    \caption{{\bf Predictors of approximability of biological networks.} Pairwise Spearman correlation between the first-, second- and third-order mean approximation errors and various network properties across the 110 published biological networks, ordered by the mean correlation. $<\cdot>$ denotes the mean, $p =$ output bias, $K =$ number of variables, $K_e =$ effective connectivity, Cov = covariance of $p(1 - p)$ and $K$. The pairwise Spearman correlations between all shown properties are in Fig.~\ref{fig:spearman_detailed}.
}
    \label{fig:spearman_simple}
\end{figure}

To test which features of a biological network make it highly approximable, we computed Spearman correlations ($\rho$) between the mean approximation errors of the 110 biological networks and several dynamics-related properties~(Fig.~\ref{fig:spearman_simple}). Highly connected networks proved less approximable ($\rho>0.6$). This is likely due to the fact that 
a continuous Taylor approximation of order $n$ matches a Boolean function with $k\leq n$ variables perfectly everywhere. Thus, the higher the average degree, $\langle K\rangle$, of a Boolean network, the lower is the chance for perfect matches. Across the three approximation orders, the average degree was even slightly more negatively correlated with network approximability than the average effective degree, $\langle K_e\rangle$, defined in~\cite{gates2021effective}. This is somewhat surprising because the latter, which takes into account the importance of Boolean inputs, is a much stronger predictor of the dynamical robustness of a Boolean network, measured by its mean average sensitivity~\cite{manicka2022effective,kadelka2024meta}. In line with this, the strongest predictor of the mean average sensitivity of a Boolean network, $\langle K_e\rangle\langle p(1-p)\rangle$, as well as the mean average sensitivity itself were both not strongly correlated with the approximability of a Boolean network, with the correlation becoming insignificant for higher-order approximations. On the contrary, the proportion of Boolean rules in a biological network, which are nested canalizing, was fairly strongly correlated with the approximability, for all orders of approximation ($|\rho| > 0.4$). The higher this proportion the more approximable was the network. 
Canalizing rules, especially those with a low sensitivity, are typically fairly biased. In line with the result on the proportion of NCFs, more biased networks proved more approximable ($|\rho| > 0.5$). Biological Boolean rules with a higher number of inputs tend to be more biased~\cite{daniels2018criticality}. Interestingly, the covariance between $p(1-p)$ and the in-degree was the only property that became more correlated with approximability at higher approximation orders. 

Metrics that explicitly describe dynamic aspects of a Boolean network also exhibited interesting correlations with the approximability. Assuming, as in the computation of approximability~\cite{manicka2023nonlinearity}, a synchronous update of all nodes, we obtained, through simulation, for each biological network a lower bound of the number of attractors, as well as the approximate mean length of the attractors, the proportion of steady state attractors and the entropy of the basin sizes (see Methods). While the third-order approximability was not correlated with any of these metrics, networks with more attractors, a lower proportion of steady state attractors and higher entropy possessed dynamics that were less approximable at first and second order. This is surprising, since the presence of many long attractors, and concomitant high entropy, is associated with Boolean networks that operate in the chaotic regime~\cite{drossel2008random}. 

\begin{figure}
    \centering
    \includegraphics[scale=0.7]{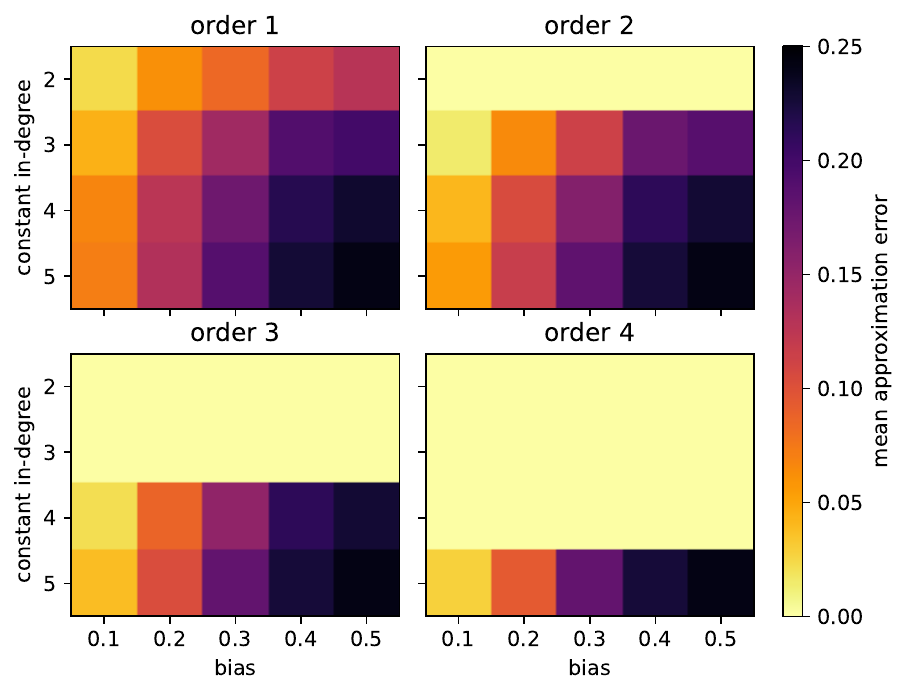}
    \caption{{\bf Effect of bias and in-degree on the approximability of the dynamics of Boolean networks.} For strongly connected $15$-node Boolean networks with a constant in-degree (y-axis) governed by random update functions generated with a certain bias (x-axis), the mean error is shown when approximating their dynamics using different order Taylor polynomials (subplots). Each cell depicts the MAE across $50$ networks, and the same networks were used to estimate the MAE using first-order to fourth-order Taylor polynomials. Results from an equivalent analysis where the functions are required to be essential in all its variables are shown in Fig.~\ref{fig:p_vs_k_nondeg}.}
    \label{fig:p_vs_k}
\end{figure}

To rule out potential confounders such as differences in network size, average degree as well as degree distribution, we considered modified $N-K$ Kauffman networks, first defined in~\cite{kauffman1969metabolic}. In these random networks of size $N$, each node has constant degree $K$. The Boolean update rule of each node is generated by drawing $2^K$ times randomly with replacement from $\{0,1\}$ with probability $1-p$ and bias $p$, respectively. We further required the wiring diagram of each network to be strongly connected since the dynamics decouple otherwise~\cite{kadelka2023modularity}. Networks with a higher absolute bias exhibited more approximable dynamics~(Fig.~\ref{fig:p_vs_k}). Moreover, sparse networks (i.e., with low in-degree) were on average more approximable. Interestingly, the MAE did not always decrease as the approximation order increased. For unbiased networks with high in-degree (e.g., $K=5$, $p=0.5$), the MAE was very close to the maximally observed value of 0.25, even when using forth-order Taylor approximations. 

Since a Boolean function with $K$ inputs is perfectly matched everywhere by a continuous Taylor approximation of order $K$, the MAE values were zero in these cases. If only $J<K$ of the inputs of a Boolean function are essential (a Boolean input is non-essential if a change in this input never changes the output of the function. For example, $f(x,y) = x$ has a non-essential input $y$), then the $J$th order Taylor approximation already provides a perfect match. To rule out a potentially confounding effect created by perfect matches, we required, in a sensitivity analysis, all update rules to be non-degenerated, i.e., to contain only essential variables~(Fig.~\ref{fig:p_vs_k_nondeg}). Most MAE values were slightly higher, likely due to the higher effective degree. Qualitatively, the results were, however, very similar.

\begin{figure}
    \centering
    \includegraphics[scale=0.7]{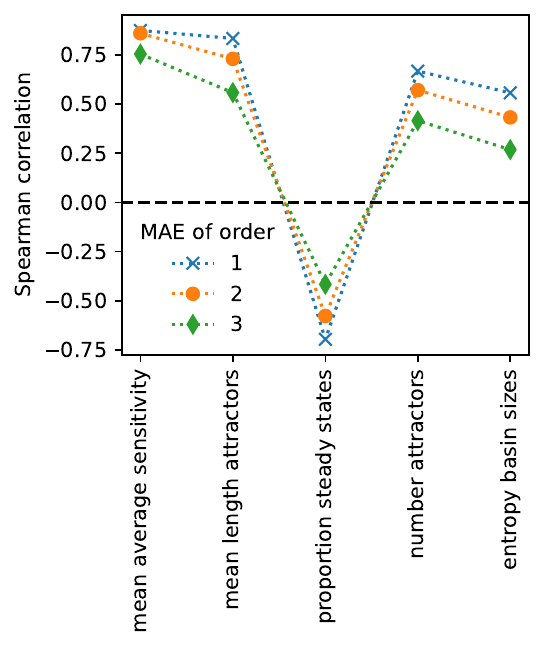}
    \caption{{\bf Predictors of approximability of random networks.} Pairwise Spearman correlation between the first-, second- and third-order mean approximation errors and network properties related explicitly to dynamics, across 2000 random strongly-connected Boolean networks with fixed degree $K\in\{2,3,4,5\}$ and bias $p\in\{0.1,0.2,0.3,0.4,0.5\}$ (100 for each combination). 
}
    \label{fig:spearman_random}
\end{figure}

Combining all 2000 random networks (100 each for combinations of constant in-degree $K\in\{2,3,4,5\}$ and bias $p\in\{0.1,0.2,0.3,0.4,0.5\}$), we computed, as before, the Spearman correlation between MAE values and metrics that explicitly describe network dynamics. The dynamical robustness of a network, measured by the mean average sensitivity, was strongly positively correlated with first-, second- and third-order MAE values ($\rho>0.75$; Fig.~\ref{fig:spearman_random}). Given that the average sensitivity of random Kauffman networks is $2Kp(1-p)$~\cite{shmulevich2004activities}, this agrees qualitatively with the results from Fig.~\ref{fig:p_vs_k}. Also in line is the finding that random networks are more approximable if they have few and short attractors, a high proportion of steady states, and low entropy in the distribution of the basin sizes. These four properties characterize networks that operate mostly in the ordered and critical dynamical regime. As observed for the biological networks, the correlations were consistently weaker when considering higher-order approximations. Note however that, by design of the computational experiment, $25\%$ ($50\%$) of the networks perfectly match their second-order (third-order) approximation, which certainly contributed to weaker correlations.

\begin{figure}
    \centering
\includegraphics[width=\textwidth]{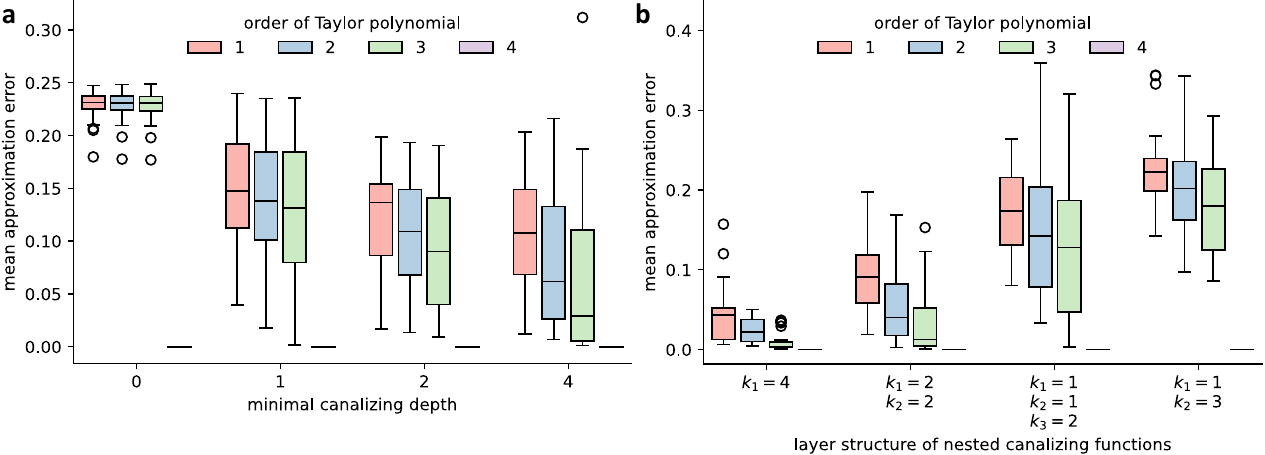}
    \caption{{\bf The approximability of Boolean network dynamics depends on canalization.} Each boxplot shows the distribution of the mean approximation error for $50$ strongly connected $N=15$-node Boolean networks with a fixed in-degree of $K=4$ and a variable degree of canalization, characterized by (a) the minimal canalizing depth of each update rule (x-axis). In (b), all functions are nested canalizing (i.e., have canalizing depth 4) but the canalizing layer structure differs.  The order of the Taylor polynomial used for the approximation is depicted by color. Fourth-order Taylor polynomials match the functions perfectly, the mean approximation error is 0. Each box extends across the interquartile range (IQR), whiskers extend to the lowest data
point still within 1.5 IQR of the lower quartile, and the highest data point still within 1.5 IQR of the upper
quartile, and black circles show outliers.}
    \label{fig:canalization}
\end{figure}

To study the effect of canalization on the nonlinearity of regulation in more detail, we modified the random networks such that the update rules were restricted to specific classes of functions. First, we compared the approximability of random networks governed by 4-variable functions with different minimal canalizing depth (see Methods). While networks without required canalization were hardly approximable (MAE $\approx 0.25$), the restriction to canalizing update rules gave rise to more approximable dynamics~(Fig.~\ref{fig:canalization}a). Functions with a higher canalizing depth are however on average also less sensitive~\cite{kadelka2017influence} and exhibit a higher absolute bias. 

While the canalizing depth provides a crude measure of the amount of canalization in a Boolean function, more detailed information is contained in the canalizing layer structure~\cite{he2016stratification,kadelka2017influence,dimitrova2022revealing}. To investigate this, we compared the approximability of random networks, each governed entirely by 4-variable NCFs but with different layer structure. Networks governed by NCFs with layer structure $k_1=4$, e.g., an AND-NOT function $x_1\wedge \bar x_2 \wedge  \bar x_3\wedge x_4$, are highly approximable~(Fig.~\ref{fig:canalization}b). On the other hand, networks governed by NCFs with layer structure $k_1=1,k_2=3$, e.g., functions such as $x_1 \vee (x_2 \wedge x_3 \wedge x_4)$, are much less approximable. Again, as the approximability of these networks decreases, the sensitivity of the underlying NCFs increases and the absolute bias decreases~\cite{kadelka2017influence}. 





\section*{Conclusion}
The idea of a probabilistic generalization of Boolean logic dates back all the way to George Boole~\cite{boole2012studies}. In this manuscript, we study in depth a recent implementation of this idea: using continuous Taylor approximations of Boolean functions to approximate the dynamics of a Boolean network. We show that the high approximability of biological networks, first postulated in~\cite{manicka2023nonlinearity}, can be almost entirely explained by the abundance of canalization in biological networks. We conjecture that the remaining higher approximability of biological networks is due to the reported increased occurrence of insensitive canalizing rules in biological networks~\cite{kadelka2024meta}. Through a computational analysis of random networks, we show that the dynamical robustness of a network strongly influences its approximability: Networks with low mean average sensitivity, operating in the ordered and critical dynamical regime and characterized by few and short attractors, possess generally more approximable dynamics. In line with this, networks governed by canalizing or even nested canalizing functions that are highly biased and insensitive to perturbations proved more approximable.

Fully disentangling the relative contribution of the related properties canalization, bias, and sensitivity on approximability constitutes one of several open questions. Moreover, it remains to be investigated how well non-perfect continuous approximations of Boolean networks perform in the context of predicting control targets or specific dynamical features. A more technical question is whether Boolean functions that can be well approximated by low-order continuous extensions give rise to more approximable Boolean networks.



\section*{Methods}
\subsection*{Boolean networks}
A \emph{Boolean network} $F$ in variables $x_1, \ldots, x_n$ can be viewed as a function on binary strings of length $n$, which can be described coordinate-wise by $n$ \emph{Boolean update functions} $f_i : \{0,1\}^n \to \{0,1\}$. 
Every Boolean network defines a canonical map, where the functions are synchronously updated:
\[
F: \{0,1\}^n\to \{0,1\}^n,\ F(x_1,\dots,x_n)=(f_1(x),\dots,f_n(x)).
\]
In this paper, we only consider this canonical map, i.e., we only consider \emph{synchronously updated Boolean networks}.

While possible, most update functions in a Boolean network do not depend on all $n$ variables. The \emph{wiring diagram} describes the dependencies. It contains $n$ nodes, corresponding to the $x_i$, and a directed edge from $x_i$ to $x_j$ if $f_j$ depends on $x_i$ (that is, if $f_j(x_1,\ldots,x_i=0,\ldots,x_n) \neq f_j(x_1,\ldots,x_i=1,\ldots,x_n)$ for at least some $(x_1,\ldots,x_{i-1},x_{i+1},\ldots,x_n)\in\{0,1\}^{n-1}$). If $f_j$ depends on $x_i$, $x_i$ is a \emph{essential} variable. Otherwise, it is \emph{non-essential}. From the wiring diagram, the degree of each node can be derived. 

\subsection*{Metrics describing Boolean network dynamics}
A second graph associated with a synchronously updated Boolean network $F$, the \emph{state space}, contains as nodes the $2^n$ binary strings and a directed edge from $x\in\{0,1\}^n$ to $y\in\{0,1\}^n$ if $F(x) = y$. Each connected component of the state space corresponds to a basin of attraction, consisting of a directed loop, the \emph{attractor}, as well as trees feeding into the attractor. Attractors can be steady states (also known as fixed points) or limit cycles. Due to its finite size, all states in a Boolean network eventually transition to an attractor. Every attractor in a biological network model typically corresponds to a distinct phenotype~\cite{schwab2020concepts}. 

Since the number of nodes, $n$, in the investigated biological Boolean network models differs from 3 to 302, some of the state spaces are huge (size $2^n$). We therefore used the following procedure to approximate several dynamics-related metrics. For each biological network $F$, we randomly picked 1000 different initial values $x_0\in\{0,1\}^n$. For each $x_0$, we synchronously updated $F$ until a repeated state was reached, indicating the arrival at an attractor. The number of updates between first and second transition to the repeated state corresponds to the length of the attractor. This process yields a non-empty list of attractors $\{A_1,\ldots,A_s\}$ of length $\{L_1,\ldots,L_s\}$ with corresponding basin sizes $\{B_1,\ldots,B_s\}$. We used $\frac 1s \sum_{i} L_i$ as the approximate mean length of the attractor and $\frac 1s \sum_{i} 1(L_i=1)$ as the approximate proportion of steady state attractors. We considered an alternative version of these two measures, weighted by the relative basin sizes (that is, $\frac 1{1000} \sum_{i} B_i L_i$ and $\frac 1{1000} \sum_{i} B_i 1(L_i=1)$). Since the alternative versions differed barely from the respective base versions (Spearman correlations of $\rho > 0.95$ across the 110 investigated biological networks), we decided to only use the base versions in the analysis. We approximated the entropy of the basin sizes as 
$$-\frac 1{1000} \sum_i \ln(\frac{B_i}{1000})B_i \in [0,\infty)$$

Finally, we used $s$ as the lower bound of the number of attractors. In a network with many attractors, we almost certainly fail to discover all attractors when starting from only 1000 random states. However, all attractors with a large basin size are discovered with high probability.

For the random Boolean networks of fixed size $n=15$, analyzed in Figs.~\ref{fig:p_vs_k}, \ref{fig:spearman_random} \ref{fig:canalization}, \ref{fig:p_vs_k_nondeg}, we computed the entire state space. All dynamics-related metrics, including the number of network attractors, are therefore exact in these analyses.

\subsection*{Continuous extensions of Boolean functions}


To compute the approximability of Boolean networks, we use the same approach as in~\cite{manicka2023nonlinearity}. We start by defining continuous extensions of Boolean functions. 
Any Boolean function $f:\{0,1\}^n\to\{0,1\}$ is defined in the corners of the $n$-dimensional hypercube, $\{0,1\}^n$, and can be extended to the entire hypercube $[0,1]^n$ by defining a function $\hat{f}:[0,1]^n\to[0,1]$ such that $\hat{f}(x)=f(x)$ for all $x\in\{0,1\}^n$. Specifically, we employ a probabilistic generalization of Boolean logic, already introduced by George Booole~\cite{boole2012studies}. We 
consider random variables $X_i:\{0,1\}\to[0,1]$ with Bernoulli distributions and set $p_i=\text{Prob}(X_i=1)$. Let $X=X_1\times\cdots\times X_n$ be the product of random variables. 
Then, we define
\[
\hat{f}(p_1,\dots,p_n) = \sum_{\substack{x\in X:\\f(x)=1}}\prod_{i=1}^n\hat{p}_i
\]
where 
\[
\hat{p}_i=\begin{cases}p_i & \text{if}\ x_i=1,\\
1-p_i & \text{if}\ x_i=0.\end{cases}
\]
By this definition, $\hat{f}:[0,1]^n\to[0,1]$ is a continuous function that satisfies $\hat{f}(x)=f(x)$ for all $x\in\{0,1\}^n$ \cite{manicka2023nonlinearity}. 

\subsection*{Taylor polynomials of Boolean functions}\label{sec:order_approx}
Since $\hat{f}$ is a continuous-variable function, we can consider different orders of approximation for $\hat{f}$ using its Taylor expansion.
As described in~\cite{manicka2023nonlinearity}, $\hat{f}$ is a square-free polynomial and its Taylor expansion is finite. More specifically, the $n^{th}$ order approximation will match $\hat{f}$ perfectly, and if only $m<n$ inputs of $f$ are essential, then the $m^{th}$ order approximation already matches $\hat{f}$ perfectly.

For a given $\alpha = (\alpha_1,\dots,\alpha_n) \in \{0,1\}^{n}$ and $x\in [0,1]^n$, we define
\begin{align*}
|\alpha| &= \alpha_1+\cdots+\alpha_n,\\
x^{\alpha} &= x_1^{\alpha_1}x_2^{\alpha_2}\cdots x_n^{\alpha_n},\\
\partial^\alpha \hat f &= \partial_1^{\alpha_1} \partial_2^{\alpha_2}\cdots\partial_n^{\alpha_n} \hat f = \frac{\partial^{\mid\alpha\mid} \hat f}{\partial_1^{\alpha_1} \partial_2^{\alpha_2}\cdots\partial_n^{\alpha_n}},
\end{align*}
with the convention that $\partial_i^{0} \hat f \equiv \hat f$. For $p\in[0,1]^n$, we have
\begin{equation}\label{eq:TaylorDec}
     \hat{f}(x) = \sum_{\alpha\in\{0,1\}^n}\frac{\partial^\alpha \hat{f}(p)}{|\alpha|!}(x-p)^\alpha=
       \hat{f}(p)+\sum_{\substack{\alpha\in\{0,1\}^n\\0<|\alpha|\leq n}}\frac{\partial^\alpha \hat{f}(p)}{|\alpha|!}(x-p)^\alpha.    
\end{equation}

If $p=(\frac 12,\ldots,\frac 12)$, which represents the unbiased selection of each variable, then $\hat{f}(p)$ equals the output bias of $f$, as shown in \cite{manicka2023nonlinearity}. The Taylor decomposition yields different approximations of a Boolean function by restricting the sum in Equation~\ref{eq:TaylorDec} to $\alpha$ with $|\alpha|\leq m \leq n$. The Taylor polynomial of order $m$ is given by
\begin{equation}\label{eq:TaylorPolynomial}
         \hat{f}^{(m)}(x) = \sum_{\substack{\alpha\in\{0,1\}^n\\|\alpha|\leq m}}\frac{\partial^\alpha \hat{f}(p)}{|\alpha|!}(x-p)^\alpha
\end{equation}

\subsection*{Approximability of a Boolean network by continuous extensions}


Let $F=(f_1,\cdots,f_n):\{0,1\}^n\to\{0,1\}^n$ be a Boolean network. We define the $m^{th}$ order approximation of $F$ to be 
\[\hat{F}^{(m)} =\Big(\max(0,\min(1,\hat{f}_1^{(m)})),\ldots,\max(0,\min(1,\hat{f}_n^{(m)}))\Big):[0,1]^n\to[0,1]^n,\] where the update functions of 
$\hat{F}^{(m)}$ are the $m^{th}$ order Taylor approximations of the update functions of $F$, $\hat{f}_i^{(m)}$ as defined in Equation~\ref{eq:TaylorPolynomial}, rescaled to the interval $[0,1]$.

With this, we can define the mean approximation error (MAE) as the mean squared error between the long-term state of the Boolean network and the long-term state of its continuous approximation. That is,
\begin{equation}
    \text{MAE}(F,m) = \frac 1{2^n}\sum_{x_0\in \{0,1\}^n}{\mid\mid F^\infty(x_0)-\hat{F}^{(m),\infty}(x_0)\mid\mid^2}
\end{equation}
where $F^\infty(x_0)$ and $\hat{F}^{(m),\infty}(x_0)$ describe the long-term state of the Boolean network $F$ and its $m^{th}$ order approximation, respectively. In practice, we approximated the MAE, using the Python library boolion~\cite{manicka2023nonlinearity}, by updating both $F$ and $\hat F^{(m)}$ synchronously 25 times and using 1000 random initial values.

\subsection*{Canalization}
This study employs several mathematical concepts related to canalization. By~\cite{kauffman1974large}, a Boolean function $f(x_1,\ldots,x_n): \{0,1\}^n \to \{0,1\}$ is canalizing if there exists a canalizing variable $x_i$, a canalizing input $a\in\{0,1\}$ and a canalized output $b\in\{0,1\}$ such that 
\begin{equation}\label{eq:canal}
f(x_1,\ldots,x_n) = \begin{cases}b & \text{if } x_i = a,\\g(x_1,\ldots,x_{i-1},x_{i+1},\ldots,x_n)\not\equiv b & \text{otherwise.}\end{cases}
\end{equation}
If the subfunction $g$ is also canalizing, then $f$ is $2$-canalizing, etc. More generally, $f$ is \emph{$k$-canalizing}, where $1 \leq k \leq n$, with respect to the permutation $\sigma \in \mathcal{S}_n$, inputs $a_1,\ldots,a_k$, and outputs $b_1,\ldots,b_k$ if
\begin{equation}\label{eq:k_canal}
f(x_{1},\ldots,x_{n})=
\left\{\begin{array}[c]{ll}
b_{1} & x_{\sigma(1)} = a_1,\\
b_{2} & x_{\sigma(1)} \neq a_1, x_{\sigma(2)} = a_2,\\
b_{3} & x_{\sigma(1)} \neq a_1, x_{\sigma(2)} \neq a_2, x_{\sigma(3)} = a_3,\\
\vdots  & \vdots\\
b_{k} & x_{\sigma(1)} \neq a_1,\ldots,x_{\sigma(k-1)}\neq a_{k-1}, x_{\sigma(k)} = a_k,\\
f_C\not\equiv b_k & x_{\sigma(1)} \neq a_1,\ldots,x_{\sigma(k-1)}\neq a_{k-1}, x_{\sigma(k)} \neq a_k.
\end{array}\right.\end{equation} 
Here, $f_C = f_C(x_{\sigma(k+1)},\ldots,x_{\sigma(n)})$ is the \emph{core function}, a Boolean function on $n-k$ variables. When $f_C$ is not canalizing, then the integer $k$ is the \emph{canalizing depth} of $f$~\cite{layne2012nested}. If $k=n$ (i.e., if all variables are become eventually canalizing), then $f$ is a \emph{nested canalizing function} (NCF)~\cite{kauffman2003random}.
By~\cite{he2016stratification}, every nonzero Boolean function $f(x_1,\ldots,x_n)$ can be uniquely written as 
\begin{equation*}
      f(x_1,\ldots,x_n) = M_1(M_2(\cdots (M_{r-1}(M_rp_C + 1) + 1)\cdots)+ 1)+ q,  
\end{equation*}
where each $M_i = \prod_{j=1}^{k_i} (x_{i_j} + a_{i_j})$ is a non-constant extended monomial, $p_C$ is the \emph{core polynomial} of $f$, and $k = \sum_{i=1}^r k_i$ is the canalizing depth. Each $x_i$ appears in exactly one of $\{M_1,\ldots,M_r,p_C\}$. The \emph{layer structure} of $f$ is the vector $(k_1,k_2,\ldots,k_r)$ and describes the number of variables in each layer $M_i$~\cite{kadelka2017influence, dimitrova2022revealing}.

\subsection*{Random null models of biological networks}
We compared biological Boolean network models to three ensembles of null models that matched different characteristics of the biological networks, as shown in Fig.~\ref{fig:bio_networks_approximability}. All null models matched the in-degree of the biological networks. Null models 1 and 3 matched, in addition, the bias of each biological update rule, while null models 2 and 3 matched the canalizing depth. 

Let $F = (f_1,\ldots,f_n)$ be a biological Boolean network model. For each $f_i$, we first simplified the function to only include essential variables, yielding $\tilde f_i :  \{0,1\}^k\to \{0,1\}$, where $k$ is the number of essential variables, i.e., the in-degree. While this step was omitted in~\cite{manicka2023nonlinearity}, it appears important for an unbiased comparison given that close to $2\%$ of regulators in biological networks are non-essential~\cite{kadelka2024meta}. We then computed the number of ones in the truth table of $\tilde f_i$, denoted $q$, and $\tilde f_i$'s canalizing depth $d$, following~\cite{dimitrova2022revealing}. 

To obtain a random Boolean function $g$ (for null model 1) with the same bias as $\tilde f_i$ and arbitrary canalizing depth, we simply selected a random subset $\Omega \subseteq \{0,1\}^k$ of size $|\Omega| = q$, and set 
\[g(x) = \begin{cases}
    1 & \text{if}\ x \in \Omega\\
    0 & \text{if}\ x \not\in \Omega. 
\end{cases}\]

To obtain a random Boolean function $g$ (for null model 2) with exact canalizing depth $d$ and arbitrary bias, we randomly selected $d$ out of $\tilde f_i$'s $k$ essential variables, arranged them in a random order, and randomly selected for each of the $d$ variables a canalizing input value $a\in\{0,1\}$ and a canalized output value $b\in\{0,1\}$ (see Equations~\ref{eq:canal}, \ref{eq:k_canal}). Finally, we randomly selected a core function $g_C: \{0,1\}^{k-d}\to \{0,1\}$, ensuring that $g_C$ depends on all $k-d$ variables and that $g_C$ is not canalizing, by repeating this random selection process until both conditions were met. We then filled the truth table of $g$, as outlined in Equation~\ref{eq:k_canal}. This entire procedure has already been implemented in the Python library canalizing\_function\_toolbox, published along with~\cite{kadelka2024meta}.

To obtain a random Boolean function $g$ (for null model 3) with the same bias as $\tilde f_i$ and the same canalizing depth $d$, we followed the same procedure as for null model 2, with two exceptions. First, we did not randomly select the canalized output values $b_1,\ldots,b_d$ but instead used the canalized output values of $\tilde f_i$. Otherwise, it is impossible to obtain the same bias. Second, we randomly selected a core function $g_C$ of $g$ that has the same number of ones as the core function of $\tilde f_i$ (following the same approach as for null model 1).

\subsection*{Random Boolean networks}
To generate a random Boolean network $F=(f_1,\ldots,f_N)$ (modified $N-K$ Kauffman network), we first generated a random  directed graph of $N$ nodes (the wiring diagram), where each node has $K$ incoming edges. We ensured the graph is simple (i.e., does not contain self-edges/auto-regulations). We further ensured the graph is strongly connected since the dynamics decouple otherwise~\cite{kadelka2023modularity}.

\begin{figure}
    \centering
    \includegraphics[scale=0.7]{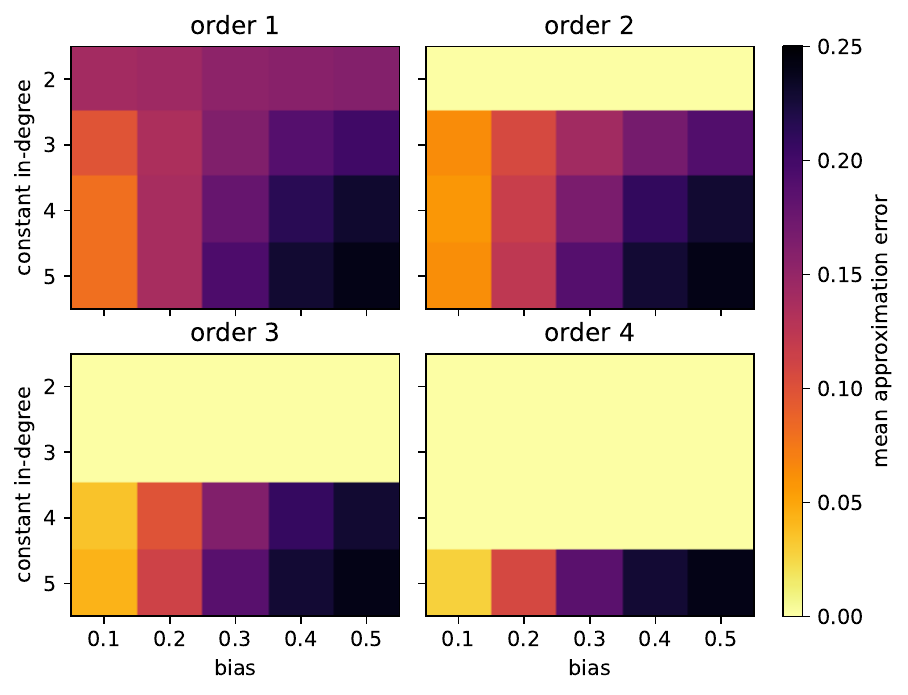}
    \caption{{\bf Effect of bias and in-degree on the approximability of the dynamics of non-degenerated Boolean networks.} For strongly connected $15$-node Boolean networks with a constant in-degree (y-axis) governed by random non-degenerated update functions generated with a certain bias (x-axis), the mean error is shown when approximating their dynamics using different order Taylor polynomials (subplots). Each cell depicts the mean approximation error across $50$ networks, and the same networks were used to estimate the mean approximation error using first-order to fourth-order Taylor polynomials. Results from an equivalent analysis where the functions are allowed to contain non-essential variables are shown in Fig.~\ref{fig:p_vs_k}.}
    \label{fig:p_vs_k_nondeg}
\end{figure}

To obtain the random Boolean update rules $f_1,\ldots,f_N$, we randomly selected, for the networks analyzed in Figs.~\ref{fig:p_vs_k}, \ref{fig:spearman_random}, any Boolean function $g: \{0,1\}^K\to \{0,1\}$. In a sensitivity analysis, reported in Fig.~\ref{fig:p_vs_k_nondeg}, we ensured that $g$ is non-degenerated, i.e., that all variables of $g$ are essential, by repeating the random selection until this condition was met. For the random Boolean networks with constant degree 4 and minimal canalizing depth $d\in\{0,1,2,4\}$, analyzed in Fig.~\ref{fig:canalization}a, we followed a very similar procedure as for null model 2 (see above), with one exception: We allowed the core function to be canalizing so that the realized canalizing depth may be larger than $d$. For the random nested canalizing Boolean networks with constant degree 4 and different layer structure, analyzed in Fig.~\ref{fig:canalization}b, we followed again a very similar procedure as for null model 2, with the exception that the layer structure determines the canalized output values,  $b_1,\ldots,b_4$~\cite{kadelka2017influence}.

\subsection*{DATA AVAILABILITY}
\cite{kadelka2024meta} contains standardized update rules of the 110 investigated published, expert-curated Boolean biological network models. 

\subsection*{CODE AVAILABILITY}
Original code to compute the approximability of Boolean networks, published along with~\cite{manicka2023nonlinearity}, is available at \href{https://gitlab.com/smanicka/boolion}{https://gitlab.com/smanicka/boolion}.

Code to analyze the 110 investigated published, expert-curated Boolean biological network models, as well as the Python library canalizing\_function\_toolbox, published along with~\cite{kadelka2024meta}, is available at \href{https://github.com/ckadelka/DesignPrinciplesGeneNetworks}{https://github.com/ckadelka/DesignPrinciplesGeneNetworks}.

New code underlying the analyses described in this manuscript is available at \href{https://github.com/ckadelka/ApproximabilityBooleanNetworks}{https://github.com/ckadelka/ApproximabilityBooleanNetworks}.

\bibliography{references}

\subsection*{ACKNOWLEDGEMENTS}
C.K. and D.M. were both partially supported by travel grants from the Simons Foundation (grant numbers 712537 and 850896, respectively).

The authors thank Iowa State University for making high-performance computing freely available to C.K.  

\subsection*{AUTHOR CONTRIBUTIONS}
Conceptualization: C.K.; Methodology: C.K. and D.M.; Software \& Visualization: C.K.; Formal analysis: C.K; Writing - Original Draft: C.K. and D.M.; Writing - Review \& Editing: C.K. and D.M.

\subsection*{COMPETING INTERESTS}
The authors declare no competing interests.

\clearpage

\setcounter{page}{1}
\beginsupplement

\backmatter

\bmhead{Supplementary information}

\begin{figure}[b!]
    \centering
    \includegraphics[width=\textwidth]{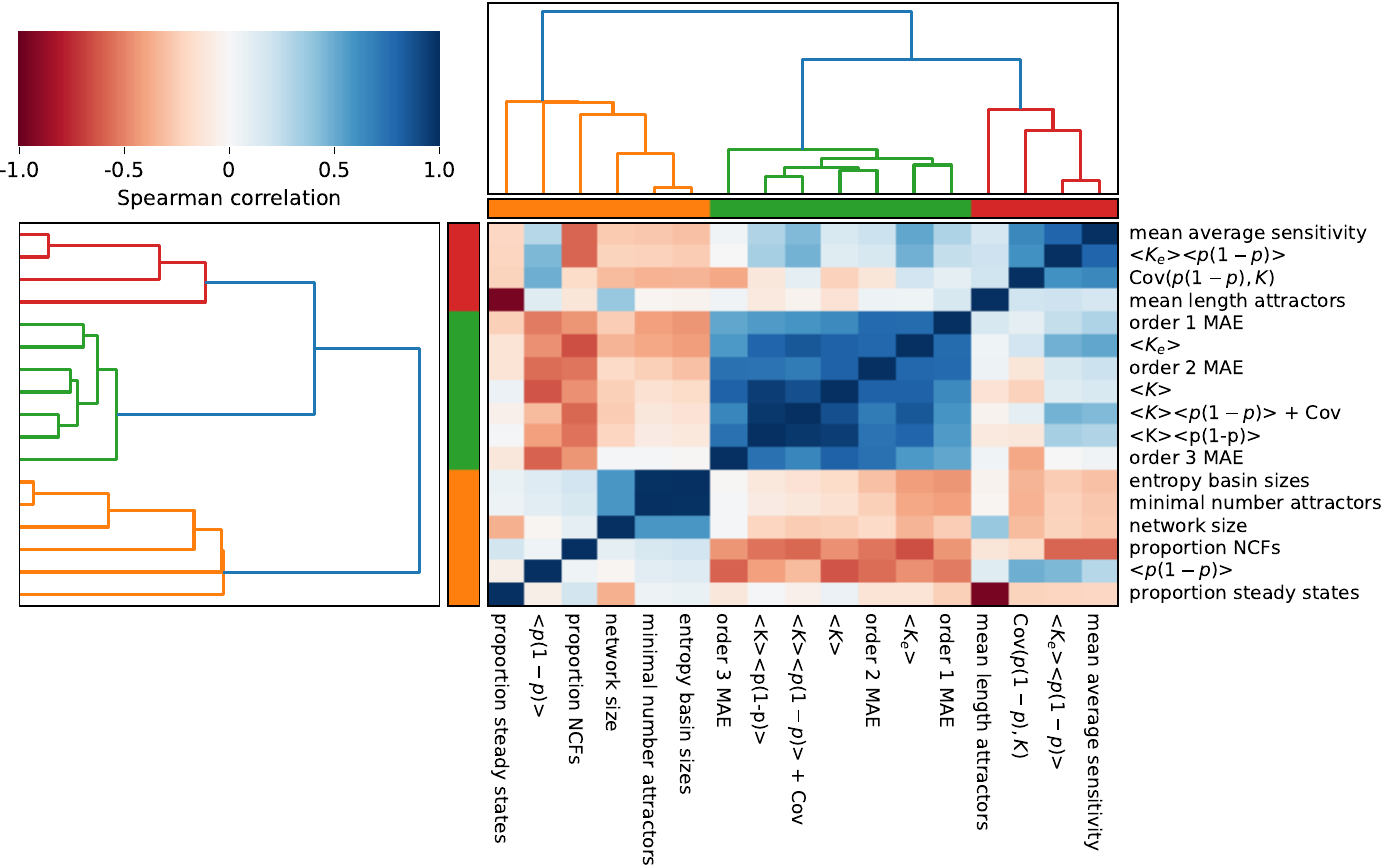}
    \caption{{\bf Pairwise Spearman correlation between properties of the 110 published biological networks.} Clusters are defined using average linkage hierarchical clustering and Euclidean distance. $<\cdot>$ denotes the mean, $p =$ output bias, $K =$ number of variables, $K_e =$ effective connectivity, Cov = covariance of $p(1 - p)$ and $K$.
}
    \label{fig:spearman_detailed}
\end{figure}

\end{document}